\documentclass[conference]{IEEEtran}

\IEEEoverridecommandlockouts
 \pdfoutput=1
\usepackage{graphicx}

\graphicspath{{./figures/}}
\usepackage{caption}
\usepackage{subcaption}
\usepackage{float}

\usepackage{amsmath} 
\usepackage{amssymb}  
\usepackage{amsfonts}
\usepackage{cite}
\usepackage{url}
\usepackage{algorithm}
\usepackage[noend]{algpseudocode}
\usepackage{color}
\usepackage{array}

\usepackage{fancyhdr}
\title{Distribution Grid Admittance Estimation with Limited Non-Synchronized Measurements
	\thanks{$*$ Xia Miao was with Siemens Corporate Technology while executing this research.}
	\thanks{Accepted to the IEEE PES GM 2019. \textcopyright  ~2019 IEEE. }
}

\author{\IEEEauthorblockN{ Xia Miao$^{*}$, Marija Ili\'{c}}
	\IEEEauthorblockA{\textit{LIDS, Massachusetts Institute of Technology}\\
		 \{xmiao,ilic\}@mit.edu}
	\and
	\IEEEauthorblockN{ Xiaofan Wu, Ulrich M\"{u}nz}
	\IEEEauthorblockA{\textit{Siemens Corporate Technology}, Princeton, NJ 08540 \\
		\{xiaofan.wu,ulrich.muenz\}@siemens.com}
   	
   }
\begin{document}

\maketitle

\begin{abstract}
 In this paper we propose a method for estimating radial distribution grid admittance matrix using a limited number of measurement devices. Neither synchronized three-phase measurements nor phasor measurements are required.  After making several practical assumptions, the  method  estimates even  impedances of lines  which have no local measurement devices installed. The computational  complexity of the proposed method is low, and this makes it possible to use for  on-line applications. Effectiveness of the proposed method is  tested using  data from a real-world distribution grid in Vienna, Austria.
\end{abstract}

\section{Introduction and motivation}
 Many countries worldwide  are pursuing the quest for  sustainable, economic and livable environment,  smart cities in particular.  The concepts of smart cities are fundamentally based on harnessing artifical intelligence (AI)  technologies and the ongoing electric power grid modernization. Seestadt Aspern, a smart city district in Vienna (Austria) \cite{ascr}, is one such example. Transforming cities into smart cities is based on  the integration of distributed energy sources (DERs) and advanced cyber (sensing, communication and control) infrastructures \cite{IWC}. Also, pro-active participation of customers is  also being enabled by embedding smart automation at the end-users level. All these innovations at the grid end users side require modernization of  low-voltage (distribution) grids, as they are  emerging as the key  enablers of  smarter and more sustainable electricity services.

Modernization of distribution power systems required by these  profound changes of end users  needs  offers new opportunities but also raises significant technical and business challenges.  To start with, one of the most basic technical challenges is the lack of accurate grid data.  This is needed because reliable and efficient  operation of a power system requires accurate state estimation and control algorithms which can not be done without good knowledge of electric power grid itself.   Recent research highlights  that errors in grid admittance parameters could lead to   significant loss of efficiency in system operation  and even cause instability\cite{Survey2000}. This is a difficult problem because the information  about the power grid is often outdated due to the limited visibility and observability in large city networks with millions of nodes.
The challenge can be tackled by beginning to rely on massive data one can collect using  fast sensors, such as the Phasor Measurement Units (PMUs). Recent progress in PMU    technologies and their rapidly decreasing cost can provide a means of collecting and  utilizing  time-series data to improve the accuracy of the grid admittance. Based on these technologies  single-line/network impedance estimation methods have been proposed for transmission grids\cite{Arafeh1979,Han2016}. However, it is realistic to assume that one cannot deploy such sensors and process all that data on all  grid lines.  To overcome this problem,  \cite{Chertkov2018,Yu2018, Low2016, MiaoACC} propose estimation methods for the case when only a limited number of sensors is  available. However, above methods all need PMU measurements from partial or all nodes, since synchronized measurements and phasor measurements are required. Although the price of PMU has dropped, it is still unrealistic and financially demanding to install PMUs (or micro-PMUs) everywhere in distribution grids. In addition, these methods assume that the system is balanced, which is not always the case  in distribution grids. The lines in distribution grids are short and are usually a mix of overhead lines and underground cables, and this  further introduces  numerical problems with grid admittance estimation. It is therefore conjectured in this paper that  estimation of  admittance in large complex distribution grids  without synchronized measurements and phasor measurements remains a challenging task. The main requirement is that any effective estimation method should be robust and should have reasonable computation requirements.

This paper concerns this basic challenge of estimating grid admittance with limited number of  conventional (non-PMU)  sensors.    In this paper we mainly focus on  admittance estimation challenges in radial distribution grids.  The proposed method is derived assuming that the network topology is known and that the non-synchronized measurements are only installed at the  limited number of nodes. Unlike other entirely data-driven approaches, our proposed method uses physical models.  A  statistical learning process utilizes this  physical model. As a result, our method is capable of providing good estimation on both  balanced and unbalanced lines.  Also an approximate estimate  of parameters is computed  for  lines where  no measurement devices  are installed. It is emphasized that no phasor measurement is required.

The rest of the paper is organized as follows: Section \ref{Sec:ProblemFormulation} provides sensor information and problem formulation. Section \ref{Sec:EstimationMethod} introduces the proposed method. Section \ref{Sec:Simulation} demonstrates using simulations  the effectiveness of the proposed method on a real-world system. And Section \ref{Sec:Conclusion} concludes the paper.

\section{Problem formulation}
\label{Sec:ProblemFormulation}
\subsection{Available sensor type and installation}
In Seestadt Aspern inexpensive measurement devices,  called Grid Monitoring Devices (GMD) \cite{GMD},  are already installed. Compared with existing PMUs,  a single GMD costs less than \$200 and  can provide  measurement  within $1\%$ accuracy.  As a trade-off, GMDs can only provide non-synchronized three-phase real power, reactive power and voltage magnitude every 2.5 minutes. Voltage phase information is not available.

The sensor installation strategy adopted in Seestadt Aspern, GMDs a has been to only  install GMDs  at one end of a single line, i.e., not to have measurements of both sending and receiving flows and voltages.  Also, loads are not measured. Although smart meters have been installed at the end-users level, these data are not available for the gird estimation task due to home privacy concerns.
\subsection{Distribution line impedance approximation in practice}
For a variety of single and multi-core cables in distribution grids, their impedances are normally approximated using the formula given in the standard IEC 60909. For example, BICC Electric Cables Handbook gives a formula for inductance as\cite{BICCBook}:
\begin{equation}
\label{Eqn:LApprox}
	L = (K + 0.2 \ln\frac{2S}{d})10^{-6}
\end{equation}
where $L$ is cable inductance ($H/m$); $K$ is conductor formation constant; $S$ is axial spacing between conductors within a cable/in trefoil/flat formulation conductors ($mm$); $d$ is conductor diameters ($mm$).

Notice that $L$ includes both self and mutual inductance. Consequently, phases of cables can be regarded as decoupled from each other, with $L$ of each phase.
\subsection{Problem formulation}
Instead of using Eqn.\eqref{Eqn:LApprox},   a method is derived to  estimate line impedances online utilizing GMD measurements. This distribution grid admittance matrix estimation problem  is  posed as follows:
\begin{itemize}
	\item Given: historical non-synchronized three-phase measurements ($P$, $Q$ and $V$) of a limited number of buses;  a known grid incident matrix $A$
	\item Objective: estimate the impedance $L$ of each line comprising of the radial distribution grid
\end{itemize}

\section{Proposed hybrid data-physics estimation method}
\label{Sec:EstimationMethod}
\begin{figure}[htp]
	\centering 
		\includegraphics[width=0.4\textwidth]{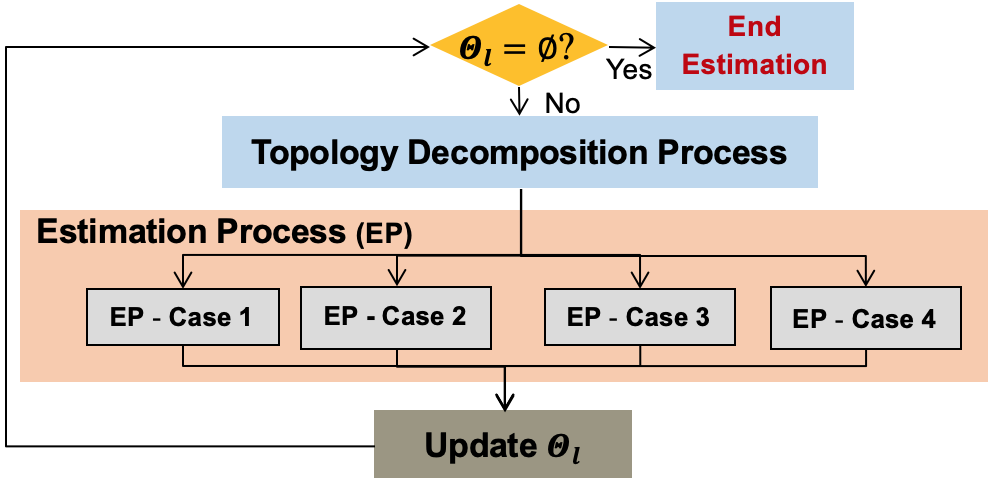}
		\caption{Structure of the proposed grid admittance estimation}
		\label{fig:DL_Structure}
\end{figure}

In this section, we propose a hybrid data-physics estimation method \cite{DLpatent} for solving the problem posed in Section \ref{Sec:ProblemFormulation}.  The method consists of a topology decomposition process and a hybrid data-physics estimation process, shown in Fig.\ref{fig:DL_Structure}. The topology decomposition process aims to break a complicated network into a few basic elements. Then, the admittance estimation is achieved by the composition of estimation procedures of basic elements (four cases in Fig.\ref{fig:DL_Structure}).
\subsection{Network topology decomposition}
\subsubsection{Basic line elements}
 Given a line, let node $i$ represent its parent (sending) node. Thus, the power flows from node $i$  to node $j$ (child node). Based on the location of GMDs, one could have four different line elements. Note that it is impossible to have real and reactive power at both sending (node $i$) and receiving end (node $j$).  GMD location and corresponding measurements of each basic element are listed in Table I. According to the available measurements, the detectability of each line element is further concluded and summarized.
\begin{figure}[htp]
	\centering 
	\scalebox{0.3}{\includegraphics{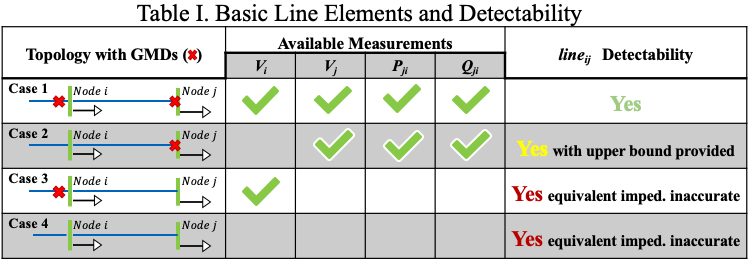}}
	\label{Table:BasicLine}
\end{figure}

The difference between Case 2 and Case 3 is: we know the real and reactive power flow received at node $j$ along $line_{ij}$ in Case 2, while we only know the total power injection at node $i$ in Case 3. Because it is possible to have multiple branches connected to node $i$. Therefore, in Case 3 and Case 4, we only provide equivalent line impedance estimation which includes possible local loads. This will be discussed in detail in Section \ref{Sec:Method}.
\subsubsection{Network topology decomposition algorithm}
Conceptually speaking, the proposed decomposition algorithm first starts with a measured node. Then, the algorithm explores its connecting lines and compares each line with  four basic elements. We first define the following notations: $\Omega_m$ is the set of nodes with GMD installed; $\Omega_N$ is the set of nodes without GMD; $\Theta_l$ is the line set; $line_{ij}$ is the line between node $i$ and node $j$; $\emptyset$ is the empty set.

The algorithm is given in Algorithm 1.

After executing Algorithm \ref{Alg:decomposition}, the system will be decomposed into basic elements listed in Table I. Different estimation processes for each basic element therefore can be invoked, which will be discussed next.
 \begin{algorithm} [htp]
 	\caption{Network Topology Decomposition Algorithm}\label{Alg:decomposition}
 	\begin{algorithmic}[1]
 		\Require  $\Omega_m$, $\Omega_N$, $\Theta_l$ (topology and sensor information)
 		\State \textbf{Start:} Let $\Omega_1 = \Omega_m$, $\Omega_2 = \Omega_N$
 		\While{$\Theta_l \neq \emptyset$}
 		\If{$\Omega_1 \neq \emptyset$}
 		\State Choose a node $j \in \Omega_1$ (bottom up direction, i.e.  child nodes first)
 		\State Remove node $j$ from $\Omega_1 ~( \Omega_1 = \Omega_1 \backslash node~j)$
 		\State Check if connecting $line_{ij} \in \Theta_l$: \textbf{No$\rightarrow$break IF}
 		\State Check if node $i\in\Omega_m$: \textbf{Yes}$\rightarrow$case 1; \textbf{No}$\rightarrow$case 2
 		\Else
 		\State Choose a node $j \in \Omega_2$ (bottom up direction)
 		\State Remove node $j$ from $\Omega_2 ~( \Omega_2 = \Omega_2 \backslash node~j)$
 		\State Check if connecting $line_{ij} \in \Theta_l$: \textbf{No$\rightarrow$break IF}
 		\State Check if node $i\in\Omega_m$: \textbf{Yes}$\rightarrow$case 3; \textbf{No}$\rightarrow$case 4
 		\EndIf
 		\State Remove $line_{ij}$ from $\Theta_l$
 		\EndWhile
 	\end{algorithmic}
 \end{algorithm}
\subsection{Hybrid data-physics estimation algorithms}
\label{Sec:Method}
Observe that distribution grid cables are usually short and GMD measurements are non-synchronized over 2.5 min period. It is reasonable to assume that measured data describes the steady-state behavior of distribution grids and the following assumption is hold.

\textbf{Assumption 1.} \textit{Voltage angle difference between two neighboring nodes is small during normal operation.}

Unlike entirely  data-driven approaches, two important physics and statistics laws form the basis of our method: Ohm's law and Law of Large Numbers \cite{LargeNumber}. Ohm’s law provides the relation between physical variables, while the Law of Large Numbers provides an efficient way of interpreting the physical meaning of the data. In the following part, we provide a sketch of derivation behind the proposed algorithms. One phase of $line_{ij}$ is used to illustrate the concept. It should be mentioned that, for each case, small modifications might be added. Notations used throughout this section are given (polar coordinate expression) as:
\begin{table}[htp]
	\centering
	\begin{tabular}{|c|c|c|c|}
		\hline
	$V_i$  & voltage magnitude of node $i$ & $I_{ij}$ & current magnitude \\
	\hline
	$\theta_i$  & voltage angle of node $i$ & $\varphi_{ij}$& current angle \\
	\hline
	$Z_{ij} $& magnitude of line $ij$ impedance &$\Phi_i$ & power factor at node $i$ \\
	\hline
	$\delta_{ij}$ &angle of line $ij$ impedance&&\\
	\hline
	\end{tabular}
\end{table}

Without loss of generality, we assume that node $i$ is the reference node of $line_{ij}$, thus has relative 0 voltage angle. Applying Ohm's law we have:
\begin{equation}
\label{eqn:1}
	(V_i\angle 0-V_j\angle \theta_j)/{Z_{ij}\angle \delta_{ij} }= I_{ij}\angle \varphi_{ij}
\end{equation}
Recall the power factor definition:
\begin{equation}
\label{eqn:2}
	\Phi_j = \theta_j -\varphi_{ij}  \quad\quad \cos\Phi_j = P_{ji}/{\sqrt{P_{ji}^2+Q_{ji}^2}}
\end{equation}
Substituting Eqn.\eqref{eqn:2} into Eqn.\eqref{eqn:1}, we obtain:
\begin{equation}
\label{eqn:3}
	V_i\angle 0-V_j\angle \theta_j = Z_{ij}I_{ij}\angle(\theta_j-\Phi_j+\delta_{ij})
\end{equation}
Thus, two equality constraints can be obtained by matching the magnitude and angle of both sides of Eqn.\eqref{eqn:3}:
\begin{align}
\label{eqn:4}
	(V_i-V_j\cos\theta_j)^2 + (-V_j\sin\theta_j)^2 & = Z_{ij}^2I_{ij}^2 \\
	\label{eqn:5}
	-V_j\sin\theta_j/(V_i-V_j\cos\theta_j) = \tan&(\theta_j - \Phi_j + \delta_{ij})
\end{align}

It can be seen that Ohm's law provides relations between measured variables and unknown variables. Next, we will show that the Law of Large Number  and Assumption 1 can be used to link the measurement data with Eqn.\eqref{eqn:4} and Eqn.\eqref{eqn:5}.

From statistics point of view, all the measured data can be regarded as random variables satisfying certain distributions. In particular, as Assumption 1 implies that $|\theta_j|\approx 0$, it is reasonable to assume that the voltage angle of measured node $\theta_j$ satisfies a distribution with zero mean. To estimate $Z_{ij}$ and $\delta_{ij}$ , the method of moments is used \cite{Moment}. Since $Z_{ij}$ and $\delta_{ij}$ do not change much over time, we thus are interested in the first order approximation from the measured data. Therefore, the physics and data can be linked through Eqn.\eqref{eqn:4} and Eqn.\eqref{eqn:5} as:
\begin{align}
 Z_{ij}^2E[I_{ij}^2] = E[(V_i-V_j&\cos\theta_j)^2 + (-V_j\sin\theta_j)^2] \notag\\
E[-V_j\sin\theta_j/(V_i-V_j\cos\theta_j) ]&= E[\tan(\theta_j - \Phi_j + \delta_{ij})] \notag
\end{align}
where $I_{ij}^2 = (P_{ji}^2+Q_{ji}^2)/{V_j^2}$ and $E[*]$ denotes the expectation of variable $(*)$.

Under Assumption 1 ($E[\theta_j]=0$) above two equations can be further simplified as:
\begin{align}
\label{eqn:6}
Z_{ij}^2E[I_{ij}^2] = E[(V_i-V_j&\cos\theta_j)^2] \\
\label{eqn:7}
 E[\tan(- \Phi_j + \delta_{ij})] = 0
\end{align}
According to the LLN, the expectation in above equations can be approximated by the mean of measured data. Before introducing specific algorithms for each case, we first define the feasible region of voltage magnitude $V_{min}$ and $V_{max}$. $V_{min} = 0.95V_{nominal}$, $V_{max} = 1.05V_{nominal}$, where $V_{nominal}$ is the rated nominal voltage magnitude.
\subsubsection{Case 1}
The available measurements are three-phase real/reactive power received at node $j$:
 \begin{equation}
 	 P = (P_{ji,a}[1], ..., P_{ji,c}[T]) \quad  Q = (Q_{ji,a}[1], ..., Q_{ji,c}[T])  \notag
 \end{equation}
 three-phase voltage magnitude of node $i$ and node $j$:
  \begin{equation}
  V_i = (V_{i,a}[1], ...,V_{i,c}[T]) \quad  V_j = (V_{j,a}[1], ..., V_{j,c}[T])  \notag
  \end{equation}
  See Algorithm \ref{Alg:case1} for detail.

\subsubsection{Case 2}
GMD is only installed at node $j$. Thus, the available measurements are three-phase real/reactive power received at node $j$ ($P/Q$) and three-phase voltage magnitude of node $j$ ($V_j$). In normal operation, voltage should satisfy the feasibility requirement. This indicates that voltage magnitude of node $i$ ($V_i$) can vary between $V_{min}$ and $V_{max}$. Estimation method is listed in Algorithm \ref{Alg:case2}.
   \begin{algorithm} [htp]
   	\caption{Estimation algorithm for case 1}
   	\begin{algorithmic}[1]
   		\Require  Three-phase measurements: $P$, $Q$, $V_i$, $V_j$
   		\State \textbf{Calculate:} the expectation of $V_i$, $V_j$, $I_{ij}$ and $\Phi_j$\\
   		$\Phi_j = acos\frac{|P|}{\sqrt{P^2+Q^2}} \rightarrow E[\Phi_j] = mean(\Phi_j)$\\
   		$I_{ij} = sign(-P)\frac{\sqrt{P^2+Q^2}}{V_j} \rightarrow E[I_{ij}] = mean(I_{ij})$\\
   		$E[V_i] = mean(V_i)\quad ~E[V_j] = mean(V_j)$
   		\State \textbf{Estimate impedance magnitude using Eqn.\eqref{eqn:6}:} \\
   		$\qquad \qquad \qquad Z_{ij} = \frac{E[V_i]-E[V_j]}{E[I_{ij}]}$
   		\State \textbf{Estimate impedance angle using Eqn.\eqref{eqn:7}:} \\
   		$\qquad \qquad \qquad \delta_{ij} = E[\Phi_j]$
   		\Ensure $Z_{ij}$, $\delta_{ij}$
   	\end{algorithmic}
   	\label{Alg:case1}
   \end{algorithm}

\begin{algorithm} [ht]
	\caption{Estimation algorithm for case 2}
	\begin{algorithmic}[2]
		\Require  Three-phase measurements: $P$, $Q$, $V_i$, $V_j$
		\State \textbf{Calculate:} $E[V_j]$, $E[I_{ij}]$ and $E[\Phi_j]$ (same as in case 1)
		\State \textbf{Estimate the upper bound of impedance magnitude:}
		\If{$E[I_{ij}]>0$}
		\State Upper bound: $\bar Z_{ij} = \frac{V_{max}-E[V_j]}{E[I_{ij}]}$
		\ElsIf{$E[I_{ij}]<0$}
		\State Upper bound: $\bar Z_{ij} = \frac{V_{min}-E[V_j]}{E[I_{ij}]}$
		\Else
		\State Upper bound: $\bar Z_{ij} = + \infty$
		\EndIf
		\State \textbf{Estimate impedance angle using Eqn.\eqref{eqn:7}:} \\
		$\qquad \qquad \qquad \delta_{ij} = E[\Phi_j]$
		\Ensure $\bar Z_{ij}$, $\underline{Z}_{ij} = 0$ and $\delta_{ij}$
	\end{algorithmic}
	\label{Alg:case2}
\end{algorithm}
\subsubsection{Case 3} GMD is installed at node $i$. Besides voltage magnitude at node $i$ ($V_i$), only three-phase total real/reactive power injection are available:
$	 P_i = (P_{i,a}[1], ..., P_{i,c}[T])$ and $Q_i = (Q_{i,a}[1], ..., Q_{i,c}[T])$.
Note that power flow along $line_{ij}$ is unknown. We therefore assume that real and reactive power are equally shared by all case 3 type lines connected to node $i$. The proposed estimation method is given in Algorithm \ref{Alg:case3}.
 \begin{algorithm} [h]
 	\caption{Estimation algorithm for case 3}
 	\begin{algorithmic}[1]
 		\Require $P_i$, $Q_i$, $V_i$ and $N_{branch} =1$
 		\State \textbf{Calculate:} $(\underline{P}_j, \underline{Q}_j) \leftarrow$ LowerBound 1( $node ~j$);
 		\State $(\underline{P}_i, \underline{Q}_i) \leftarrow$ LowerBound 2(node $i$, node $j$, $N_{branch}$);
 		\State \textbf{Calculate:} the approximated real/reactive power injection to the $line_{ij}$ using:\\ $\qquad \qquad P_{ij} = \frac{E[P_i]-\underline{P}_i -\underline{P}_j}{N_{branch}} \quad Q_{ij} = \frac{E[Q_i]-\underline{Q}_i -\underline{Q}_j}{N_{branch}}$
 		\State \textbf{Estimate:} Resistance and reactance of equivalent impedance as: $R_{ij} = \frac{E[V_i]^2}{P_{ij}} \quad X_{ij} = \frac{E[V_i]^2}{Q_{ij}}$
 		\Ensure $Z_{ij} = \sqrt{R_{ij}^2+X_{ij}^2} \quad \delta_{ij} = atan(\frac{X_{ij}}{R_{ij}})$
 		\Function{LowerBound 1}{node $j$}
 		\For {$m \in$ child nodes of node $i$ }
 		\If{$line_{jm}$ has been estimated}
 		\State $P_{jm} = E[I_{jm}]^2Z_{jm}\cos(\delta_{jm})-E[P_{mj}]$
 		\State $Q_{jm} = E[I_{jm}]^2Z_{jm}\sin(\delta_{jm})-E[Q_{mj}]$
 		\Else
 		\State $P_{jm} = 0 \qquad Q_{jm} = 0$
 		\EndIf
 		\EndFor
 		\State \textbf{Return:}  $\underline{P}_j=\sum_{m}P_{jm} \quad \underline{Q}_j=\sum_{m}Q_{jm}$
 		\EndFunction
 		\Function{LowerBound 2}{node $i$, node $j$, $N_{branch}$}
 		\For {$n \in$ child nodes of node $i ~\&\&~n\neq j$ }
 		\If{$line_{in}$ has been estimated}
 		\State $P_{in} = E[I_{in}]^2Z_{in}\cos(\delta_{in})-E[P_{ni}]$
 		\State $Q_{in} = E[I_{in}]^2Z_{in}\sin(\delta_{in})-E[Q_{ni}]$
 		\Else
 		\State $N_{branch} = N_{branch}+1$
 		\State  $[\underline{P}_{in}, \underline{Q}_{in}] \gets$ LowerBound 1 (node $n$)
 		\EndIf
 		\EndFor
 		\State \textbf{Return:} $\underline{P}_i = \sum_{n}P_{in}$, $\underline{Q}_i = \sum_{n}Q_{in}$ and $N_{branch}$
 		\EndFunction
 	\end{algorithmic}
 	\label{Alg:case3}
 \end{algorithm}
\subsubsection{Case 4}
We do not have GMD and thus no measurements are available related to the line. However, we can still approximate the impedance by making the following assumption. The proposed method for case 4 is shown in Algorithm \ref{Alg:case4}.

\textbf{Assumption 2:} Given a node $i$, we assume that all case 4 type lines connected to node $i$ have same impedance $Z$ and $\delta$.

\begin{algorithm} [h]
	\caption{Estimation algorithm for case 4}
	\begin{algorithmic}[1]
		\State let $N_{branch} =1$
		\State $(Z, \delta, N_{branch}) \gets Case4(node~i, N_{branch}) $
		\State \textbf{Calculate:} $Z_{ij} = Z/N_{branch} \quad \delta_{ij} = \delta$
		\State Assign $Z_{ij},\delta_{ij}$ to all case 4 type lines in Function Case4
		\Ensure $Z_{ij} = \sqrt{R_{ij}^2+X_{ij}^2} \quad \delta_{ij} = atan(\frac{X_{ij}}{R_{ij}})$
		\Function{Case4}{node $i$, $N_{branch}$}
		\State \textbf{Find:} parent node of node $i\rightarrow$ node $k$
		\If{node $k$ has GMD}
		\State ($Z_{ki}, \delta_{ki} )$ $\gets$ \textbf{call:} Algorithm \ref{Alg:case3} ($line_{ki}$)
		\Else
		\State $(Z, \delta, N_{branch}$)$\gets$ Case4(node $k$, $N_{branch}$)
		\EndIf
		\State $N_{branch} = N_{branch}+1$
		\State \textbf{Return:}  $Z_{ki}, \delta_{ki}, N_{branch}$
		\EndFunction
		\Ensure $Z_{ij} = \frac{Z}{N_{branch}} \qquad \delta_{ij} = \delta$
	\end{algorithmic}
	\label{Alg:case4}
\end{algorithm}
\subsection{Computational complexity and online implementation}
It can be seen that proposed algorithms only require basic algebraic operations and corresponding computational complexity is $O(n)$. Therefore, it is possible to implement the method online which has been validated on a subnet in Aspern smart city. Results and required data size analysis are discussed in Section \ref{Sec:Simulation}.
\section{Evaluation With Real-World System}
\label{Sec:Simulation}
\subsection{System description}
\begin{figure}[ht]
	\centering
	\includegraphics[width=0.5\textwidth]{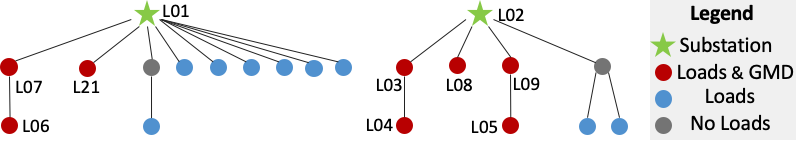}
	\caption{An exemplary distribution subnet in Aspern Smart City}
	\label{fig:example}
\end{figure}
The Seestadt Aspern distribution system has 11 radial subnets connecting to their substations in the normal condition.  Notice that all subnets can be decomposed into four basic line elements. In order to evaluate the proposed methods, a subnet in Aspern smart city is considered, whose topology with sensor installation map is shown in Fig.\ref{fig:example}.

GMD IDs are directly marked next to their installed nodes in Fig.\ref{fig:example}. One month (May 2018) measurements (around 15,000 data points) are used as input data. It is worthwhile mentioning that measurement noise and errors are unavoidable. For example,  data points are missing for certain dates. Also, we observe from measurements that not all lines are balanced.
\subsection{Estimation results discussion}
In the test, the benchmark impedance value for each line is calculated by multiplying line length and corresponding equivalent resistance and inductance value found in the data sheet. The estimation results for case 1 and case 2 type lines are listed in the Table II.
\begin{figure}[htp]
	\centering
	\includegraphics[width=0.43\textwidth]{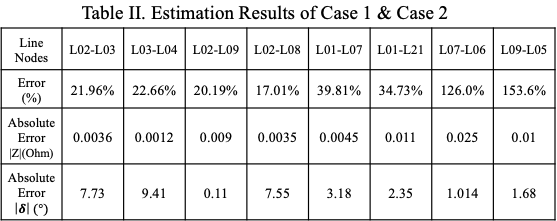}
\end{figure}

 It can be seen that the proposed algorithm managed to estimate the line impedance based on GMD data. As can be seen from Table II, the estimation results are accurate. Although the average percentage error is around 23$\%$, the absolute error of both magnitude and angle are very small. In particular, the error of the Line between L09 and L05 is the largest, at around 153$\%$. However, the absolute error of the angle is less than 1.8 degree, which is even smaller than the other lines.

 Given that real line impedance is unknown, percentage error may not be a good measure on estimation accuracy. However, a higher percentage error may indicate that the actual GMDs installation on the line L09-L05 is different from the information found in the data sheet. Or some additional devices have been installed on the line in practice which has changed the line impedance but such situation has not been reflected in the manual. This conjecture is supported by the result that the estimated impedance magnitude is higher than the benchmark value. It is worthwhile to have another GMD installed at L05 side. Therefore, more accurate line impedance could be obtained and the corresponding result can be used to validate the proposed estimation method.

For the rest case 3 and case 4 type lines, as discussed in Section \ref{Sec:EstimationMethod}, we can only provide equivalent impedances which includes line impedances and connecting local loads. Thus, our estimation results for these two type lines are close to their actual value only when local loads are small. Because of this observation, we omit case 3 and 4 estimation results for brevity. However, it should be mentioned that such a situation can be improved if additional information from local smart devices are provided, such as data from E-meters, etc.
\subsection{Sensitivity analysis}
\begin{figure}[htp]
	\centering
	\includegraphics[width=0.35\textwidth]{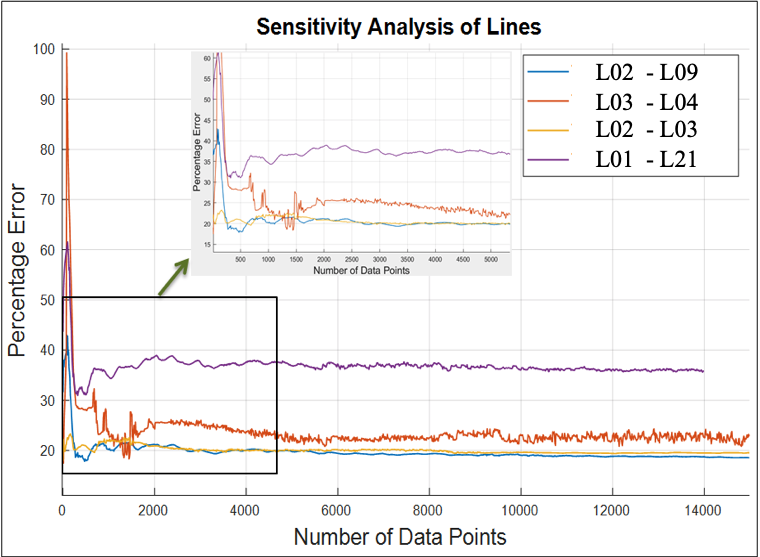}
	\caption{Sensitivity analysis of selected lines}
	\label{fig:sensitivity}
\end{figure}

We conduct sensitivity analysis on selected lines with respect to the size of data, since computation time depends on the size of utilized data points.  The result is shown in Fig.\ref{fig:sensitivity}. The estimation error tends to decrease as more data points are utilized. However, it does not improve much when more than 5000 data points are used. Notice that one day measurements is around 500 data points. Thus, in practice, there is no need to use all available data points. 5-10 days measurements can provide a reasonably good result.

\section{Conclusions}
\label{Sec:Conclusion}
In this paper we propose a novel algorithm which is able to estimate the radial distribution admittance by only using non-synchronized measurements from limited locations. More importantly, phasor measurement is not required. It should be also emphasized that the computational complexity of the proposed hybrid data-physics method is remarkably low, resulting in an easy online implementation with a small board. Furthermore, our preliminary tests, conducted on a real-world system, show that the proposed method performs well when noise and unbalanced network are considered. In the future, we plan to combine the proposed real-time estimation with the  state estimation to improve the operation efficiency.


\begin{thebibliography}{99}
\bibitem{ascr}
	https://www.ascr.at/
\bibitem{IWC}
	https://www.irena.org/publications/2018/Mar/RenewableStatistics2018
\bibitem{Survey2000}
Zarco, Pedro, et al. "Power system parameter estimation: a survey." IEEE Trans. on power systems (2000).
\bibitem{Arafeh1979}
Arafeh, Samir A., and R. Schinzinger. "Estimation algorithms for large-scale power systems." IEEE Trans. on Power Appar. and Syst. (1979)
\bibitem{Han2016}
Han, Sekyung, et al. "An automated impedance estimation method in low-voltage distribution network for coordinated voltage regulation." IEEE Transactions on Smart Grid 7.2 (2016): 1012-1020.
\bibitem{Cobreces2009}
Cobreces,et al. "Grid impedance monitoring system for distributed power generation electronic interfaces." IEEE Trans. Instru. and Meas. 
\bibitem{Yu2018}
Yu, Jiafan, et al. "PaToPa: A data-driven parameter and topology joint estimation framework in distribution grids." IEEE Transactions on Power Systems 33.4 (2018): 4335-4347.
\bibitem{Chertkov2018}
Park, Sejun, et al. "Learning with End-Users in Distribution Grids: Topology and Parameter Estimation." arXiv:1803.04812 (2018).
\bibitem{Low2016}
 Yuan, Ye, Omid Ardakanian, Steven Low, and Claire Tomlin. "On the inverse power flow problem." arXiv preprint arXiv:1610.06631 (2016).
 \bibitem{MiaoACC}
 X. Miao, et al."Multi-layered Grid Admittance Matrix Estimation for Electric Power Systems with Partial Measurements", submit ACC 2019.
 \bibitem{GMD}
 Siemens TD-3551/EMMS30: https://www.ait.ac.at/fileadmin/mc/energy
\bibitem{BICCBook}
BICC Electrical Cables Handbook. Wiley, Dec 8, 1997

\bibitem{DLpatent}
X. Miao, et al. "Distribution Grid Admittance Estimation with limited Non-Synchronized Measurements", USPTO No: 62/756,311
\bibitem{LargeNumber}
Yu.V. Prokhorov, "On the strong law of large numbers" Izv. Akad. Nauk SSSR Ser. Mat., 14 (1950) pp. 523–536
\bibitem{Moment}
Bowman, KO and Shenton, LR(1998). "Estimator: Method of Moments", Encyclopedia of Statistical Sciences, Wiley, pp 2092-2098
\end{thebibliography}
\end{document}